\nonstopmode

\documentclass[11pt]{article}

\usepackage[fleqn]{amsmath}

\usepackage{hyperref}
\hypersetup{
    colorlinks=true,
    linkcolor=blue,
    filecolor=magenta,      
    urlcolor=blue,
}

  \usepackage{geometry}
\usepackage{amsmath,amssymb,amsthm}
\usepackage{graphics,epsfig,calc}

\usepackage{latexsym,epsfig,bm,amssymb}
\usepackage{xcolor}
\usepackage{amsthm,mathrsfs}

\DeclareSymbolFont{AMSb}{U}{msb}m{n}
\DeclareSymbolFontAlphabet{\mathbb}{AMSb}

\newcommand{\HS}{{\rm HS}}

\newcommand{\cA}{{\cal A}}

\newcommand{\cD}{{\cal D}}

\newcommand{\da}{\dagger}
\newcommand{\ci}{\cite}
\newcommand{\de}{\delta}
\newcommand{{\De}}{{\Delta}}

\newcommand{\fr}{\frac}

\newcommand{\ga}{\gamma}

\newcommand{\lam}{\lambda}

\newcommand{  \om}{  \omega}

\newcommand{ \ov}{ \overline}

\newcommand{\si}{\sigma}

\newcommand\C{{\mathbb C}}
\newcommand\RR{{\mathbb R}}

\newcommand{\tr}{\mathop{\rm tr\,}\nolimits}

\newtheorem{theorem}{Theorem}[section]

\newtheorem{definition}[theorem]{Definition}

\newtheorem{lemma}[theorem]{Lemma}

\newtheorem{remark}[theorem]{Remark}

\newtheorem{corollary}[theorem]{Corollary}

\begin{document}
\begin{center}

{\huge  On dissipation operators of Quantum Optics}

\bigskip\bigskip\bigskip
{\large A.I. Komech$^1$ and E.A. Kopylova}\footnote{ 
 The research was funded in whole by Austrian Science Fund (FWF) 10.55776/PAT3476224.}
 \\
{\it
Institute of Mathematics of
BOKU
University, Vienna, Austria\\
}
 alexander.komech@boku.ac.at,\quad
 elena.kopylova@boku.ac.at

\smallskip

\bigskip

  \hspace{8cm} To the memory of Mikhail Shubin

\end{center}

\setcounter{page}{1}
\thispagestyle{empty}



  \begin{abstract}

  We consider  dissipation operators used in
  Quantum Optics for the description of quantum 
  spontaneous emission in the context of
 damped driven  Jaynes--Cummings equations.
The equations describe quantised one-mode Maxwell field
coupled to a  two-level 
molecule.

The nonpositivity of two basic  dissipation operators
is proved
 in the framework of the theory of the Hilbert space of the Hilbert-Schmidt  
  operators. We show that one of the operators is symmetric, while the other is not.

     \end{abstract}

  \smallskip
  

\tableofcontents

\setcounter{equation}{0}
\section{Jaynes--Cummings equation}
The
Jaynes--Cummings equation  (JCE)  is basic model of Quantum Optics. The model without damping and pumping was introduced in 1963 by 
Jaynes and Cummings
\ci{JC1963} (the survey can be found in \ci{LM2021}). 
The  damped  driven Jaynes--Cummings equations  read as  
\ci[(5.107)]{VW2006}:
\begin{equation}\label{eq1.1}
\dot \rho(t)=\cA(t)\rho(t):=-i[H(t),\rho(t)]+\ga D\rho(t),\qquad t\ge 0.
\end{equation}
Here $\rho(t)$ is the density operator of the coupled field-molecule system,  
i.e., it is  a nonnegative  Hermitian operator with the trace one, in the Hilbert space 
$X=F\otimes\C^2$, where $F$ is the separable Hilbert space
endowed with 
an orthonormal basis $|n\rangle$, $n=0,1,\dots$
and the annihilation operator $a$ and its adjoint, creation operator
$a^\dag$:
\begin{equation}\label{eq1.2} 
 a|n\rangle=\sqrt{n}|n-1\rangle,\qquad 
 a^\dag|n\rangle=\sqrt{n+1}|n+1\rangle, \qquad [a,a^\da]=1.
\end{equation}
 The Hamiltonian $H(t)$ is the sum
 $$
H(t)\!=\!H_0\!+\!pV(t),\,\,\, H_0\!=\!H_F\!+\!H_A,\,\,\, H_F\!=\!\om_c a^\da a,
\,\,\, H_A\!=\!
\fr12  \om_a\si_3,\,\,\, V(t)\!=\! \si_1(a\!+\!a^\dag)\!+\!A^e(t).
$$
 Here $H_0$ is the Hamiltonian of the free field and atom without interaction,
while $pV(t)$ is the interaction Hamiltonian,
$\om_c>0$ is the cavity resonance frequency,  
$\om_a>0$ is the molecular
frequency, $\ga>0$ and $D$ is a dissipation operator,
  $p\in\RR$ is proportional to the molecular dipole moment.
  By
   $A^e(t)$ we denote the pumping, 
$\si_1$ and $\si_3$ are the Pauli matrices
 acting on $\C^2$, so $[a,\si_k]=[a^\dag,\si_k]=0$.

\noindent{\bf Acknowledgements.} 
The authors thank S. Kuksin, M.I. Petelin, A. Shnirelman and H. Spohn
 for longterm fruitful discussions.

\setcounter{equation}{0}
\section{Dissipation operators and main results}

The dissipation operator $D$ in (\ref{eq1.1}) describes the 
quantum spontaneous emission which provides an energy
loss to balance the pumping.
The dissipation operators must satisfy the following conditions.
\smallskip

We will consider  the dissipation operators
 \ci[(5.107)]{VW2006} and \ci[(5.6.20)]{BR1997}:
      $$
      D\rho=a\rho a^\dag-\fr12 a^\dag a\rho-\fr12 \rho a^\dag a
+a^\dag\rho a-\fr12 a a^\dag\rho-\fr12 \rho a a^\dag,
  \qquad
 \De\rho=a\rho a^\dag-\fr12 a^\dag a\rho-\fr12 \rho a^\dag a.
$$
\begin{definition}
  $\HS$ is the Hilbert space of Hermitian 
    Hilbert--Schmidt operators with the inner product {\rm \ci[Thm VI.22 (c)]{RS1980}} 
\begin{equation}\label{eq2.1}
\langle\rho_1,\rho_2\rangle_\HS=\tr[\rho_1\rho_2].
\end{equation}
\end{definition}
 
 \begin{definition}
  i) $|n,s_\pm\rangle=|n\rangle\otimes s_\pm$ is the orthonormal basis in $X$, where
    $s_\pm\in \C^2$ and
$
\si_3 s_\pm=\lam(s_\pm) s_\pm$ with
$\lam(s_\pm)=\pm 1.
$
\smallskip\\
ii) $X_0$ is the space 
of finite linear combinations of the vectors $|n,s_\pm\rangle$.\smallskip\\
iii) $\cD\subset\HS$ is the subspace of finite rank Hermitian operators 
\begin{equation}\label{eq2.2}
\rho=\sum_{n,n'} \,\sum_{s,s'=s_\pm}\rho_{n,s;n',s'}
|n,s\rangle\otimes \langle n',s'|.
\end{equation}
\end{definition}
Every density operator $\rho\in\HS$
    is defined uniquely by its matrix entries
      \begin{equation}\label{eq2.3}
      \rho_{n,s;n',s'}=\langle n,s|\rho|n',s'\rangle,
      \end{equation}
      which are Hermitian $2\times2$ matrices.
The Hilbert--Schmidt norm, corresponding to  the inner product  (\ref{eq2.1}), can be written as
\begin{equation}\label{eq2.4}
\Vert\rho \Vert_\HS^2=\tr[\rho^2]=\sum_{n,n'=0}^\infty\,\sum_{s,s'=s_\pm} |\rho_{n,s;n',s'}|^2<\infty.
\end{equation}

   The operator $D$
 is a modification of
 the version $\De$.
 The version $\De$ has been introduced
 in \ci{A1973}--\ci{A1974}, and
   used
  in
 \ci[(5.6.20)]{BR1997}
  and \ci[(18)]{BJ2007}, \ci{L1976}, \ci[(4)]{SF2002}. The version 
   is symmetric in the following  two aspects: 
   for $\rho\in\cD$, 
   $\De\rho$ is the 
  selfadjoint operator  
  with  zero trace that is necessary for the dynamics of $\rho(t)$ in  selfadjoint
  operators with constant trace.
  The modification $D$
  restores the third symmetry with respect to  the interchange of $a$ and $a^\dag$.

    The main result of present paper is the following theorem.
  
    \begin{theorem}\label{th2.3}
i) Both  dissipation operators $D$ and $\De$ are
defined on the domain $\cD\subset\HS$ and 
nonpositive:
$$
\langle\rho,D\rho\rangle_\HS\le 0, \qquad\langle\rho,\De\rho\rangle_\HS\le 0,\qquad\rho\in\cD.
$$
ii) The operator  $D:\cD\to\cD$ is symmetric:
$$
\langle\rho_1,D\rho_2\rangle_\HS=\langle D\rho_1,\rho_2\rangle_\HS,\qquad\rho_1,\rho_2\in\cD,
$$
while $\De$ is not symmetric.
\smallskip\\
iii) Both operators $D,\De:\cD\to\cD$ are injective. 
\end{theorem}

    \begin{corollary}
    The dissipation operator $D$ admits the selfadjoint Friedrichs extension.
    \end{corollary}

\setcounter{equation}{0}
\section{Nonpositivity and symmetry}
Here we prove Theorem \ref{th2.3} i) and ii). First, 
let us 
prove the nonpositivity of $\De$. 
For $\rho\in\cD$,
\begin{eqnarray}\label{eq3.1}
\langle\rho,\De\rho\rangle_\HS&=&\tr\big(\rho \De\rho\big)
=\tr\Big(\rho\big(
a\rho a^\dag-\fr12 a^\dag a\rho-\fr12 \rho a^\dag a
\big)
\Big)
\nonumber\\
\nonumber\\
&=&
\tr\big(\rho a\rho a^\dag-\rho a^\dag a\rho
\big)
=\tr\big(\rho a\rho a^\dag- a^\dag a\rho^2
\big).
\end{eqnarray}
Now we use the fact that $\rho$ is a finite rank Hermitian operator (\ref{eq2.2}).
Then (\ref{eq1.2}) implies that
 the operators $\rho a\rho a^\dag$ and $a^\dag a\rho^2$
have only  finite 
number of nonzero entries (\ref{eq2.3}), so their traces 
are well-defined. 
Moreover,
$\rho$ admits a finite spectral resolution  in the orthonormal basis of its eigenvectors  $e_i\in X_\infty$:
$$
\rho=\sum_{i=1}^\nu\rho_i e_i\otimes e_i.
$$
In this basis, the entries  $\rho_{ij}=\rho_i\de_{ij}$, and 
the entries 
$a_{jk}=\langle e_j,a e_k\rangle$
and $a^\dag_{kl}=\langle e_k,a^\dag e_l\rangle$
of the operators $a$ and $a^\dag$ are well-defined.
Hence, (\ref{eq3.1}) implies,
with summation in repeated indices, 
\begin{eqnarray}\label{eq3.2}
\langle\rho,\De\rho\rangle_\HS&=&
\rho_i\de_{ij}a_{jk}\rho_k\de_{kl}
a^\dag_{li}
-a^\dag_{kl}a_{lj}\rho_j^2\de_{jk}
%
=
\rho_i a_{ik}\rho_k
a^\dag_{ki}
-a^\dag_{kl}a_{lk}\rho_k^2
\nonumber\\[1ex]
&=&
\rho_i a_{ik}\rho_k
a^\dag_{ki}
-a^\dag_{ki}a_{ik}\rho_k^2
=
a_{ik}a^\dag_{ki}(\rho_i\rho_k
-\rho_k^2
)
\nonumber\\
&=&
\fr12\Big(a_{ik}a^\dag_{ki}(\rho_i\rho_k
-\rho_k^2)+a_{ki}a^\dag_{ik}(\rho_k\rho_i
-\rho_i^2) 
)
=
-\fr12
|a_{ik}|^2(\rho_i-\rho_k)^2
\le 0
\end{eqnarray}
since $a^\dag_{ik}=\ov a_{ki}$. Now,  the nonpositivity  is proved for $\De$.

\begin{lemma} $D=\De+\De^\dag$, where $\De^\dag$ differs from $\De$ by  swapping
$a$ and $a^\dag$:
\begin{equation}\label{eq3.3}
\De^\dag\rho=a^\dag\rho a -\frac 12 \rho a^{\dag}a-\frac 12a^{\dag}a\rho,\qquad \rho\in\cD.
\end{equation}
\end{lemma}

\begin{proof}

For  $\rho_1,\rho_2\in\cD$,
 \begin{eqnarray}\nonumber
\tr [\rho_1(\De\rho_2)]
&=&\tr
\Big(
\rho_1(a\rho_2 a^\dag-\frac 12 a^{\dag}a\rho_2-\frac 12\rho_2 a^{\dag}a
)
\Big)
\\
\nonumber
&=&\tr\Big(
\rho_1a\rho_2 a^\dag-\frac 12 \rho_1a^{\dag}a\rho_2-\frac 12\rho_1\rho_2a^{\dag}a 
\Big)\\
\nonumber
&=&\tr
\Big(
a^\dag\rho_1a\rho_2 -\frac 12 \rho_1a^{\dag}a\rho_2-\frac 12a^{\dag}a\rho_1\rho_2 
\Big)\\
\nonumber
&=&\tr
\Big((a^\dag\rho_1a -\frac 12 \rho_1a^{\dag}a-\frac 12a^{\dag}a\rho_1 
)\rho_2 
\Big)
=
\tr[(\De^\dag\rho_1)\rho_2].
\end{eqnarray}
Hence,
(\ref{eq3.3}) is proved.
\end{proof}
As the corollary, the nonpositivity of $\De^\dag$ holds by the same arguments as for $\De$ just by  swapping
$a$ and $a^\dag$,
and hence, $D=\De+\De^\dag$ is also nonpositive. Moreover, $D$ is symmetric on $\cD$, while
$\De$ is not.

\begin{remark}\rm
The proof of the nonpositivity 
   essentially depends on the symmetry of  $\rho$.
 \end{remark}

\setcounter{equation}{0}
\section{Injectivity}
Let us prove Theorem \ref{th2.3} iii).
The
formula (\ref{eq3.2}) gives
\begin{equation}\label{eq4.1}
\langle\rho,\De\rho\rangle_\HS=-\fr12
|a_{ik}|^2[\rho_i-\rho_k]^2,\qquad\rho\in\cD.
\end{equation}
It implies 
that $\langle\rho,\De\rho\rangle=0$
 only if $a_{ik}=0$ for $\rho_i\ne\rho_k$. Therefore, the eigenspaces of $\rho$ corresponding to  the eigenvalues $\rho_i$ are invariant with respect to $a$ and
$a^\dag$. However, there is only one such eigenspace, the entire
Hilbert space  $X$. Accordingly, 
all eigenvalues $\rho_i$ coinside. 
Hence, all the eigenvalues 
are zeros since the zero is in spectrum 
of any density operator $\rho\in\cD$.

\begin{remark}\rm
The last step uses the fact that $\rho\in\cD$.
Alternatively, we can use the argument that
the Hilbert--Schmidt norm (\ref{eq2.4}) is infinite if all the eigenvalues $\rho_i$ are nonzero and identical.
This argument also works for general $\rho\in\HS$
if the relation (\ref{eq4.1}) holds.
However, our proof of this relation is correct only for 
$\rho\in\cD$.
\end{remark}

\end{document}